# Raman signatures of monoclinic distortion in $(Ba_{1-x}Sr_x)_3CaNb_2O_9$ complex perovskites

**Authors**

J. E. Rodrigues [1]; D. M. Bezerra [2]; R. C. Costa [3,4]; P. S. Pizani [3]; A. C. Hernandes [1]

**Affiliations**

[1] Crystal Growth and Ceramic Materials Group, São Carlos Institute of Physics, University of São Paulo, CEP 13560-970, São Carlos SP, Brazil.

[2] São Carlos Institute of Chemistry, University of São Paulo, CEP 13566-590, São Carlos SP, Brazil.

[3] Department of Physics, Federal University of São Carlos, CEP 13565-905, São Carlos SP, Brazil.

[4] Department of Environmental Engineering, Federal University of Campina Grande, CEP 58840-000, Pombal PB, Brazil.

**Corresponding author**

J. E. Rodrigues

E-mail address: rodrigues.joaoelias@gmail.com or rodrigues.joaoelias@ursa.ifsc.usp.br.

URL: http://www.ifsc.usp.br/ccmc.

**Abstract**

Octahedral tilting is most common distortion process observed in centrosymmetric perovskite compounds (ABO$_3$). Indeed, crucial physical properties of this oxide stem from the tilts of BO$_6$ rigid octahedra. In microwave ceramics with perovskite-type structure, there is a close relation between the temperature coefficient of resonant frequency and tilt system of the perovskite structure. However, in many cases, limited access facilities are needed to assign correctly the space group, including neutron scattering and transmission electron microscopy. Here, we combine the Raman scattering and group-theory calculations to probe the structural distortion in the perovskite (Ba$_{1-x}$Sr$_x$)$_3$CaNb$_2$O$_9$ solid solution, which exhibits a structural phase transition at $x \geq 0.7$, from $D_{3d}^3$ trigonal to $C_{2h}^3$ monoclinic cell. Both phases are related by an octahedral tilting distortion ($a^0b^-b^-$ in Glazer notation). Low temperature Raman spectra corroborate the group-theoretical predictions for Sr$_3$CaNb$_2$O$_9$ compound, since 36 modes detected at 25 K agree well with those 42 (25**A**$_g$ ⊕ 17**B**$_g$) predicted ones.

*Keywords*: (Ba$_{1-x}$Sr$_x$)$_3$CaNb$_2$O$_9$; octahedral tilting; Raman scattering; factor-group analysis.

# 1. Introduction

It is well-established that the compounds crystallizing in the perovskite structure ($ABO_3$) display an enormous technological and scientific importance in solid state physics and chemistry [1,2]. Due to their ability to accommodate a large variety of cations at A- and B-sites, several physical properties may emerge from this feature, including ferroelectricity, piezoelectricity, superconductivity, ionic conductivity, half-metallic conductivity, multiferroicity, and ferromagnetism [3,4]. With a correct chemical design, the parameters of each property can be tuned to improve their performance as sensors, transducers, capacitors, photovoltaic cells, electrolytes for fuel cell, magnetic memories, catalysts, or dielectric resonators [5–7]. From a structural point of view, the substitution at the A or B sites can provide two interesting phenomena in centrosymmetric perovskite structures, namely, octahedral tilting and B-site cation ordering [8,9]. In the first case, cooperative tilts of $BO_6$ rigid octahedra should occur to accommodate a smaller A-site cation in the cuboctahedral cavity [10]. In the second one, two distinct cations, B' and B'', yield an ordered pattern and then inducing a superstructure formation. As a consequence, these processes modify the overall symmetry of the crystalline structure from the simple cubic perovskite ($O_h^1$ or *Pm-3m*) [11,12].

Indeed, the mechanisms behind the chemical design optimization are usually the octahedral tilting and B-site cation ordering, being promising in the case of the dielectric resonator devices for microwave applications [13,14]. In the last decades, such devices have been applied successfully in the telecommunication fields, including millimeter-waves, base station, and mobile phones [15]. For high performance, the microwave dielectric ceramics should present a high dielectric permittivity ($\varepsilon'$), low dielectric loss (tan $\delta \sim 10^{-4}$) and near zero-temperature coefficient of resonant frequency ($\tau_f$). By using the chemical design, for instance, it is possible to improve the dielectric loss at microwave by controlling the B-site cation ordering at long-range [16]. On the other hand, a temperature-stable device can be achieved by inducing the octahedral tilting [17,18]. Both approaches are frequently employed to obtain desired microwave parameters in the 1:2 ordered perovskite, with the chemical formula $A^{2+}_3B'^{2+}B''^{5+}_2O_9$ [A = Ba, Sr, Ca; B' = Ca, Mg, Co, Zn; B'' = Ta, Nb]. In its undistorted state, such a structure adopts a trigonal unit cell belonging to the $D_{3d}^3$ space group, with an alternately distribution of the B-site cations along the $\langle 111 \rangle_c$ cubic cell direction [9].

Recently, we have reported in the $Ba_3CaNb_2O_9$ compound that the dielectric loss is related to the 1:2 ordered domain amount in the crystallite. Also, we have argued that the origin of its high $\tau_f$ value stems from the structural instabilities associated with the onset of an octahedral tilt transition [19]. In this sense, $Sr^{2+}$ is selected as the A-site substituent cation for the $Ba_3CaNb_2O_9$ perovskite. Thus, on the basis of ionic radii [20], the strontium ion may introduce distortions in the crystalline structure. In this paper, we synthesized the $(Ba_{1-x}Sr_x)_3CaNb_2O_9$ solid solution to probe possible distortion induced by the octahedral tilting [$CaO_6$ and $NbO_6$ units]. As far as we know, there are no available reports on the structural features concerning this system. Furthermore, Raman spectroscopy is applied to detect the local structure disturbance induced by symmetry lowering. As reported by many authors [21–24], the combination of Raman scattering and factor-group analysis is an important tool to investigate the structural changes in perovskites and related compounds through the optical phonons behavior at the first-Brillouin zone ($\mathbf{q} \approx 0$). This approach will be used here to assign our Raman results.

2. Experimental details

All polycrystalline samples were prepared using the conventional solid state synthesis method under air atmosphere. We aim to synthesize the $(Ba_{1-x}Sr_x)_3CaNb_2O_9$ solid solution (hereafter: BCN-SCN system). Stoichiometric amounts of $BaCO_3$ [Alfa Aesar: 99.80%], $SrCO_3$ [Sigma Aldrich: 99.90%], $CaCO_3$ [Alfa Aesar: 99.95%] and $Nb_2O_5$ [Alfa Aesar: 99.90%] were weighed and homogenized in a nylon jar containing isopropyl alcohol and zirconia cylinders for 24 h. After a dry procedure at 100 °C, the mixture was thermally treated at 1300 °C for 2 h and then grounded to obtain fine powders.

X-ray powder diffraction patterns (XRPD) were acquired by a *Rigaku* Rotaflex RU-200B diffractometer (Bragg-Brentano θ-2θ geometry; Cu-Kα radiation λ = 1.5406 Å; 50 kV and 100 mA) over a 2θ range from 10° to 100° with a step size of 0.02° [step time of 1 s]. Raman spectra were collected in a backscattering geometry by a *Jobin-Yvon* T64000 triple monochromator with a $LN_2$ cooled CCD camera and a 1800 grooves.mm$^{-1}$ holographic grating. An Olympus microscope is attached to the spectrometer and the Raman signals were excited by a 514 nm line of an argon laser (Coherent, Innova 70C). For low temperature measurements (25 up to 300 K), a helium closed cycle cryostat [with a temperature controller *LakeShore* model 330] was employed. All spectra were further corrected by the Bose-Einstein thermal factor prior to the fitting procedure using the Lorentzian line shape, including a baseline subtraction.

## 3. Glazer notation and group theory analysis

In this work, we will describe the octahedral tilting process in perovskite using a notation designed by Glazer [10]. Each tilt system depends on the $BO_6$ octahedral rotations about the axes of the so-called aristotype unit cell (an undistorted structure). Such a notation has the following representation: $a^\#b^\#c^\#$, where the letters represent the tilt magnitude about the *x*, *y* and *z* orthogonal axes, respectively. The # symbol denotes no tilt (# = 0), antiphase (# = -) and in-phase octahedral tilting (# = +) in neighboring layers. Therefore, the aristotype perovskite is designated by $a^0a^0a^0$ in Glazer notation. In the 1:2 ordering, the undistorted structure belongs to the trigonal $D_{3d}^3$ space group (*P-3m*1; #164), for which 9 ($4\mathbf{A}_{1g} \oplus 5\mathbf{E}_g$) Raman active phonons are expected, as summarized in Table 1. Particularly, we are interested in distorted structure with 1:2 ordering at the B-site derived from antiphase tilting process [9]. In Table 1, it is also depicted the site symmetries and the corresponding vibrational modes for the crystalline structures addressed here [25].

[**TABLE 1**]

## 4. Results and discussion

XRPD patterns of the $(Ba_{1-x}Sr_x)_3CaNb_2O_9$ ($x$ = 0.0, 0.3, 0.5, 0.7, 0.9, and 1.0) system are illustrated in Figure 1 (a). Within the detection limit of the X-ray diffraction technique, no phase segregation was noted in the full-range composition. The end members of the BCN-SCN system were indexed in accordance with the ICSD#162758 and ICDD#17-0174 cards, respectively, for the BCN and SCN systems. Those cards are related to a trigonal crystal structure belonging to the $D_{3d}^3$ space group (*P-3m*1; #164) with one chemical formulas per cell, in which the $Ca^{2+}$ and $Nb^{5+}$ ions feature an ordering behavior, occupying the 1*a* and 2*d* Wyckoff sites, respectively. The earlier structure is derived from the disordered cubic cell within the $O_h^1$ space group, as a consequence of rhombohedral distortion along the ‹111›$_c$ cubic direction [26], which provides the following lattice parameters for the resulting trigonal cell: $a_h \approx \sqrt{2}a_c$ and $c_h \approx \sqrt{3}a_c$ ($a_c$ is the cubic lattice constant). Once the $Sr^{2+}$ ion has a lower ionic radius than that for the $Ba^{2+}$ ion, one can expect the *d*-spacing shifting behavior. Indeed, a shifting can be noted with increasing strontium content in Figure 1 (b), resulting in a decrease of the unit cell volume, as shown in Figure 1 (c).

Some diffraction peaks at the 2θ region from 10° to 25° start to appear for $x \geq 0.7$. At the same time, an extra peak at 2θ ~ 36.5° becomes more intense, being indexed as ½(3 1 1)$_c$ reflection from the simple cubic ABO$_3$ perovskite. As reported for similar systems [27,28], a phase transition induced by the oxygen octahedra tilting may cause the appearing of this peak in the patterns of Figure 1 (a). The atomic substitution for smaller ions at the A-site leads to the crystal symmetry lowering owing to the A-O bond decrement along with the octahedral tilting, in which the BO$_6$ octahedra are tilted in order to preserve the connectivity of their corners [10,12]. Such a behavior increases the number of the diffraction peaks as a result of the symmetry lowering process [29,30]. Figure 1 (c) compiles the strontium content dependence of the cell volume and theoretical density of the (Ba$_{1-x}$Sr$_x$)$_3$CaNb$_2$O$_9$ compositions. The linear trend of both curves can be regarded a signature of the solid solution formation in the system under investigation. Also, there is a good agreement between the earlier reported densities (for BCN and SCN systems) and those obtained in this work.

[**FIGURE 1**]

Nagai *et al*. proposed that the antiphase tilting of both MgO$_6$ and TaO$_6$ octahedra changed the space group of the (Ba$_{1-x}$Sr$_x$)$_3$MgTa$_2$O$_9$ compounds from trigonal (D$_{3d}^3$) to monoclinic (C$_{2h}^6$; *A*2/*n* or *C*2/*c*; #15) [27]. Another evidence of this phase transition comes from the (2 0 2)$_h$ peak splitting into two counterparts indexed as (2-4 2) and (2 4 2) reflections from the monoclinic phase [28,29]. This behavior was not noted here for the (2 0 2)$_h$ peak [see Figure 1 (b)], although a slight increase in its intensity occurred for $x \geq 0.9$. According to Goldschmidt [31], the distortion and structural stability of the perovskite structures can be predicted using the tolerance factor (*t*). For the BCN–SCN system, the tolerance factor can be obtained as follows:

$$t = \frac{1}{\sqrt{2}} \frac{\left[(1-x)R_{Ba} + xR_{Sr}\right] + R_O}{\frac{1}{3}(R_{Ca} + 2R_{Nb}) + R_O}, \quad (1)$$

so that $R_{Ba}$, $R_{Sr}$, $R_{Ca}$, $R_{Nb}$ and $R_O$ are the ionic radii of Ba (1.61 Å; CN = 12), Sr (1.44 Å; CN = 12), Ca (1.00 Å; CN = 6), Nb (0.64 Å; CN = 6) and O (1.35 Å; CN = 2), taken from the Shannon's table [20]. The calculated $t$ value was around 0.992 for BCN, which decreases linearly until 0.935 for SCN. It means that the BCN-SCN may experience a phase transition as the $Sr^{2+}$ substitution increases. Hence, one can expect a distorted 1:2 structure for SCN rather than the undistorted $D_{3d}$ cell for BCN [29]. To investigate those structural changes in more detail, Raman scattering measurements were collected in all samples of the BCN-SCN solid solution at room temperature, as discussed next.

Raman spectra of the $(Ba_{1-x}Sr_x)_3CaNb_2O_9$ solid solution are displayed in Figure 2 (a). In the spectrum of BCN ($x = 0.0$), the peaks at 85 and 90 cm$^{-1}$ are assigned to $\mathbf{E}_g$ and $\mathbf{A}_{1g}$ external modes, respectively. The bands centered at 413 and 820 cm$^{-1}$ denote the internal modes concerning the stretching and breathing motions of NbO$_6$ octahedra. The twisting breath-mode of oxygen octahedron occurs at 355 cm$^{-1}$ with $\mathbf{E}_g$ symmetry. The peaks appearing at approximately 135, 250, 280, 610 cm$^{-1}$ can be attributed to the $D_{3d}$ trigonal structure [19]. Two remaining bands are due to the 1:1 domain boundary in the crystalline structure, as elucidated by Blasse *et al* [32]. Such bands are allowed when the crystallites exhibit the accumulation of antisite defects, defining the antiphase boundary defects [5]. It is clear that this phenomenon occurred in the full-range composition. Recently, Ma *et al.* analyzed the boundary regions of 1:2 ordered domain in Ba$_3$(Co, Zn, Mg)Nb$_2$O$_9$ perovskites using HRTEM technique, detecting an extra stabilized ordered structure in this regions [33]. In practice, it means that there are two well-defined zones in the crystallites containing 1:1 (domain boundary) and 1:2 (domain) orders at B-site. Therefore, we should expect the fingerprints of each region in the Raman spectra at room temperature.

[**FIGURE 2**]

Figure 2 (b) illustrates the effect of strontium content on the peak parameters ascribed to the 1:1 and 1:2 regions, respectively. One can note the redshift behavior of the intense oxygen breathing $\mathbf{A}_{1g}$ mode ranging from 820 ($x = 0.0$) to 811 cm$^{-1}$ ($x = 1.0$). The 1:2 peak area has a minimum value at $x = 0.5$, corresponding to a maximum one for the 1:1 peak area. Such a trend can better visualized by the contour map in Figure 3. As a whole, this phenomenon can be interpreted as follows: the Sr$^{2+}$ incorporation leads to an increasing in the domain boundary distribution until $x = 0.5$. From this value, it was noted a decreasing in this boundary region. In a recent work, we found out that there is dependence between the 1:1 peak parameters and the dielectric performance of the Ba$_3$CaNb$_2$O$_9$-based microwave ceramics. In particular, a decreasing in the 1:1 peak intensity led to an improved quality factor [19]. Azough *et al.* also reported that the high quality factor ($Q \times f$) depends on the removal of domain boundaries in single domain grains (crystallites) of Ba$_3$MgTa$_2$O$_9$ ceramics [34]. Therefore, we can predict that the Sr$^{2+}$ cations at the A-site not only modify the thermal stability (i.e $\tau_\mathbf{f}$ parameter), but also the dielectric loss at microwave.

[**FIGURE 3**]

Unlike BCN sample, there are an increased number of Raman bands in the SCN spectrum, as depicted in Figures 4(a) and 4(b). Our results are in accordance with those reported for Sr$_3$MgTa$_2$O$_9$ perovskites [30]. Dias *et al.* showed that Sr$_3$MgNb$_2$O$_9$ system can be indexed as a monoclinic cell belonging to the C$_{2h}^3$ space group (*A2/m*; #12). This assumption is based on the Howard and Stokes seminal work describing the group-subgroup relationships for $A_3B'B''_2O_9$ possible phases [9,28]. From the untitled trigonal phase ($D_{3d}^3$; $a^0a^0a^0$ in Glazer notation), five derived structures can be obtained directly from the octahedral tilting process, namely: $C_{3i}^1$ (*P*-3; #147; $a^+a^+a^+$), $C_{2h}^3$ (*C*2/*m*; #12; $a^0b^+b^+$), $C_{2h}^5$ (*P*2$_1$/*c*; #14; $a^0a^0c^+$), $C_{2h}^3$ (*A*2/*m*; #12; $a^0b^-b^-$) e $D_{3d}^4$ (*P*-3*c*1; #165; $a^-a^-a^-$). As one can see, only two of them are ascribed as an antiphase (out-of-phase) octahedral rotation with a negative superscript. An antiphase rotation usually appears in perovskites with a negative value for the temperature coefficient of resonance frequency ($\tau_f$) [35]. Such a parameter denotes the thermal stability of the dielectric resonator for applications in microwave circuitry [36]. Therefore, the monoclinic *A*2/*m* and trigonal *P*-3*c*1 phases should better describe the crystalline structure of $Sr_3CaNb_2O_9$ composition. By using the method of factor group analysis [25], we calculated the distribution of the zone-center vibrational modes in terms of the respective representations of $C_{2h}$ and $D_{3d}$ point groups, as listed in Table 1.

[**FIGURE 4**]

In this sense, 42 (25$\mathbf{A_g} \oplus$ 17$\mathbf{B_g}$) Raman active modes are expected to be detected for the monoclinic structure, while 20 (6$\mathbf{A_{1g}} \oplus$ 14$\mathbf{E_g}$) Raman active phonons are allowed for the trigonal space group. Otherwise, it was noted 29 bands in the SCN Raman spectrum at room temperature, as depicted in Figure 4. This result can eliminate the trigonal system ($a^-a^-a^-$ in Glazer notation) in the description of the SCN structure. However, low temperature Raman spectra were collected in order to solve overlapping bands from the thermal effect, and the results are shown in Figure 5. Also, the spectra were acquired in the ranging from 25 to 300 K and then corrected by the Bose-Einstein thermal factor [$n(\omega,T) + 1$] [37]. The bands in the wavenumber interval 65-515 cm$^{-1}$ are clearly more detectable, leading 36 modes at approximately 25 K. Such a result agrees more closely with the 42 ones predicted for the $C_{2h}^3$ monoclinic system ($a^0b^-b^-$ in Glazer notation), being our finding similar to that reported for $Sr_3MgNb_2O_9$ [28]. Since the internal modes of $NbO_6$ octahedron usually appears at $\tilde{\upsilon} > 550$ cm$^{-1}$, we should not expect more than two bands in this interval (modes at 613 and 812 cm$^{-1}$) [38]. The remaining band centered at 776 cm$^{-1}$ comes from the domain boundary, as previously mentioned.

[**FIGURE 5**]

According to Howard *et al* [9], the $C_{2h}^3$ monoclinic structure for the $Sr_3CaNb_2O_9$ system has three $Sr^{2+}$ at 4*i* positions, two $Ca^{2+}$ located at 2*a* and 2*d* sites, two $Nb^{5+}$ at 4*i* positions and six oxygen anions distributed at 4*i* (3$O_1$) and 8*j* (3$O_2$) Wyckoff sites. Its unit cell has the dimensions $\approx 2\sqrt{3} \times \sqrt{2} \times \sqrt{6}$ relative to the edge of the $ABO_3$ simple cubic perovskite. Otherwise, a more distorted structure is expected to crystallize when $Sr^{2+}$ are replaced by $Ca^{2+}$ cations. Indeed, Levin *et al*. concluded that the $Ca_4Nb_2O_9 \equiv Ca_3CaNb_2O_9$ perovskite can be indexed as a monoclinic cell belongs to the $C_{2h}^5$ space group ($P2_1/c$; #14; $b^-b^-c^+$ in Glazer notation) [39]. Besides the antiphase rotation, this structure has an in-phase octahedral tilting with the cell dimensions $\approx \sqrt{6} \times \sqrt{2} \times 3\sqrt{2}$ relative to the edge of the cubic perovskite. Fu *et al*. also employed the previous monoclinic unit cell to describe the crystalline structure of the $Ca_3MgB''_2O_9$ (B'' = Nb and Ta) complex perovskite ceramics [40].

## 5. Conclusions

In short, we have investigated the $(Ba_{1-x}Sr_x)_3CaNb_2O_9$ solid solution obtained by conventional solid state method. Our main goal in this work was to introduce a structural transition by replacing $Ba^{2+}$ cations with $Sr^{2+}$. A combination of Raman spectroscopy and factor group analysis was promising to detect the symmetry lowering process in Sr-rich samples ($x \geq 0.7$). Based on the Howard and Stokes work, we argued that the monoclinic structure within the $C_{2h}^3$ space group is an appropriate choice to describe the unit cell of $Sr_3CaNb_2O_9$. Such a monoclinic cell is directly obtained from the aristotype $D_{3d}^3$ trigonal cell after an octahedral tilting distortion ($a^0b^-b^-$ in Glazer notation). Indeed, low temperature Raman spectra were employed to corroborate the group-theoretical predictions and a total of 36 peaks were fitted at 25 K in comparison with those 42 ($25\mathbf{A}_g \oplus 17\mathbf{B}_g$) predicted ones. As a result, we believe that the highlight of this paper lies in the understanding of the structural distortion induced by octahedral tilting process in the $(Ba_{1-x}Sr_x)_3CaNb_2O_9$ solid solution, and then this result can encourage futures studies in similar systems.

**Acknowledgments**

Financial support from the Brazilian funding agencies CAPES [BEX 0270/16-4], CNPq [Proc. Number 573636/2008-7], INCTMN [Proc. Number 2008/57872-1], and FAPESP [Proc. Number 2013/07296- 2] is acknowledged. J. E. Rodrigues would like to thank Pr. Olivier Thomas for all support at Aix-Marseille Université.

**Figure Captions**

Figure 1:

**Crystal structure probed by XRPD technique.** (a) X-ray diffraction pattern for the $(Ba_{1-x}Sr_x)_3CaNb_2O_9$ powder sample, at room temperature, indicating the *d*-spacing shifting behavior, (b) see the $(2\ 0\ 2)_h$ peak. (c) The unit cell volume and theoretical density values are shown to illustrate their linear dependence with strontium content. Although the BCN and SCN structures were indexed in agreement with the $D_{3d}$ trigonal structure, the indices of planes are labeled following the pseudo-cubic setting (space group $O_h^5$).

Figure 2:

**Raman crystallography.** (a) Raman spectra for the $(Ba_{1-x}Sr_x)_3CaNb_2O_9$ powder sample, indicating the changes in low wavenumber region in Sr-rich samples ($x \geq 0.7$). (b) Strontium content dependence of the peaks parameters (center and normalized area) ascribed to the 1:1 and 1:2 ordered regions.

Figure 3:

**1:1 and 1:2 ordered regions.** The contour plot of the Sr content dependence of the Raman spectra in high wavenumber region. The $Sr^{2+}$ incorporation leads to an increasing in the domain boundary distribution until $x = 0.5$.

Figure 4:

**Comparison between trigonal and monoclinic phases.** Raman spectra for the $Ba_3CaNb_2O_9$ and $Sr_3CaNb_2O_9$ compounds in the wavenumber regions: (a) 65-465 cm$^{-1}$ and (b) 490-840 cm$^{-1}$, showing an increased number of bands in the strontium-based system.

Figure 5:

**Monoclinic phase confirmation.** Low temperature Raman spectra of the $Sr_3CaNb_2O_9$ composition in the regions: (a) 65-515 cm$^{-1}$ and (b) 525-875 cm$^{-1}$. A total of 36 peaks were fitted at 25 K in comparison with those 42 ($25\mathbf{A}_g \oplus 17\mathbf{B}_g$) predicted ones.

**Table Captions**

Table 1:

Factor group analysis for the crystal structures addressed in this work ($\Gamma_T = \Gamma_{Ac} \oplus \Gamma_{Si} \oplus \Gamma_{IR} \oplus \Gamma_R$).

Figure 1:

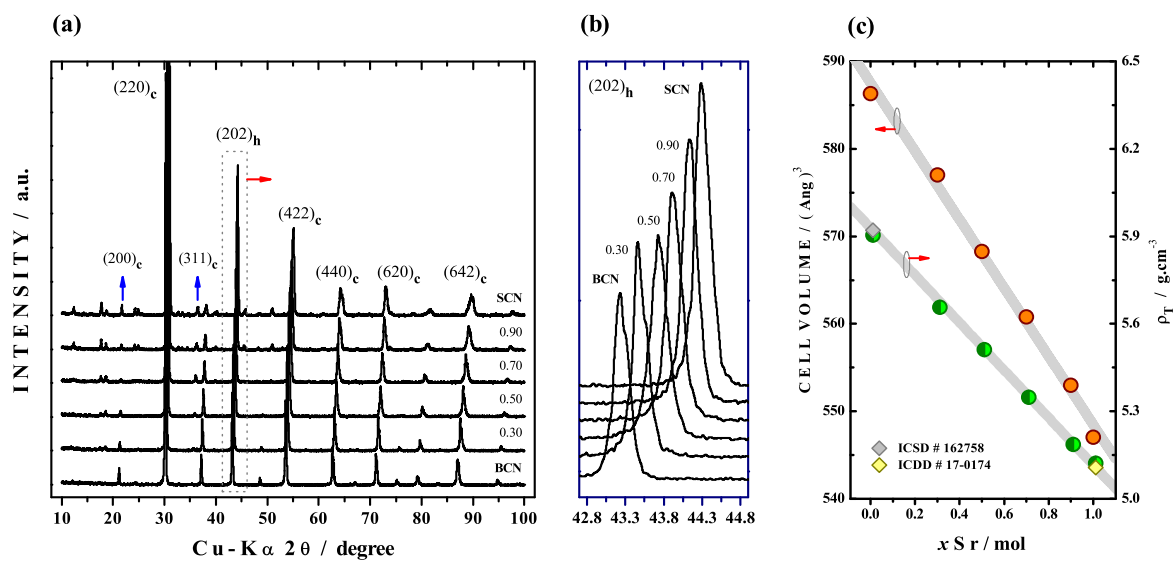

Figure 2:

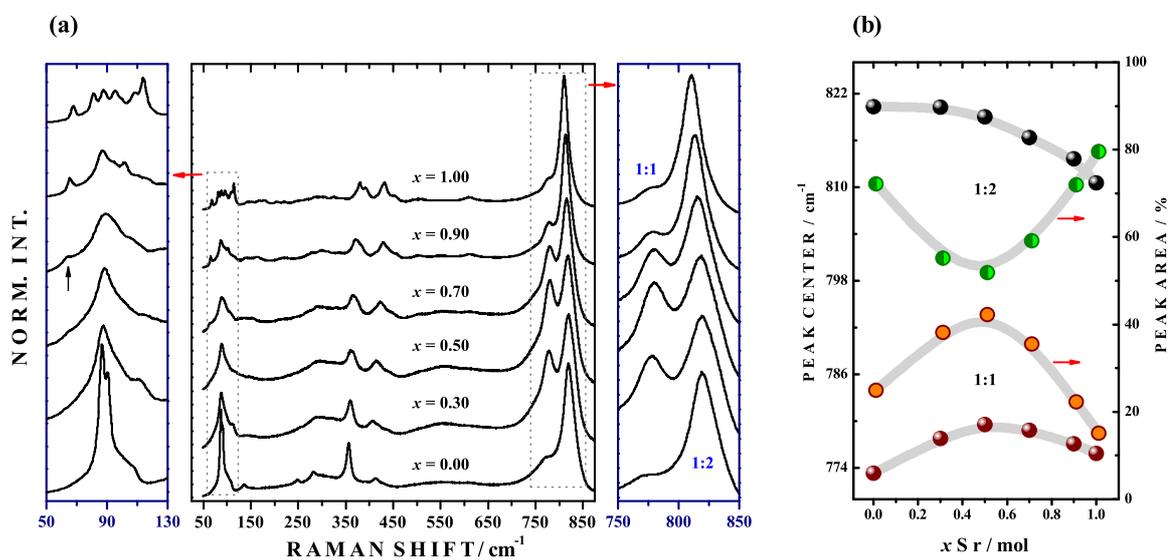

Figure 3:

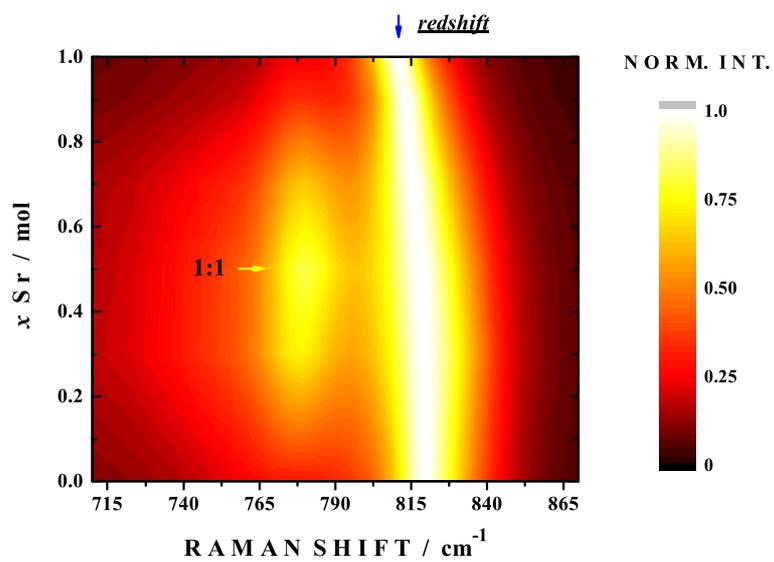

Figure 4:

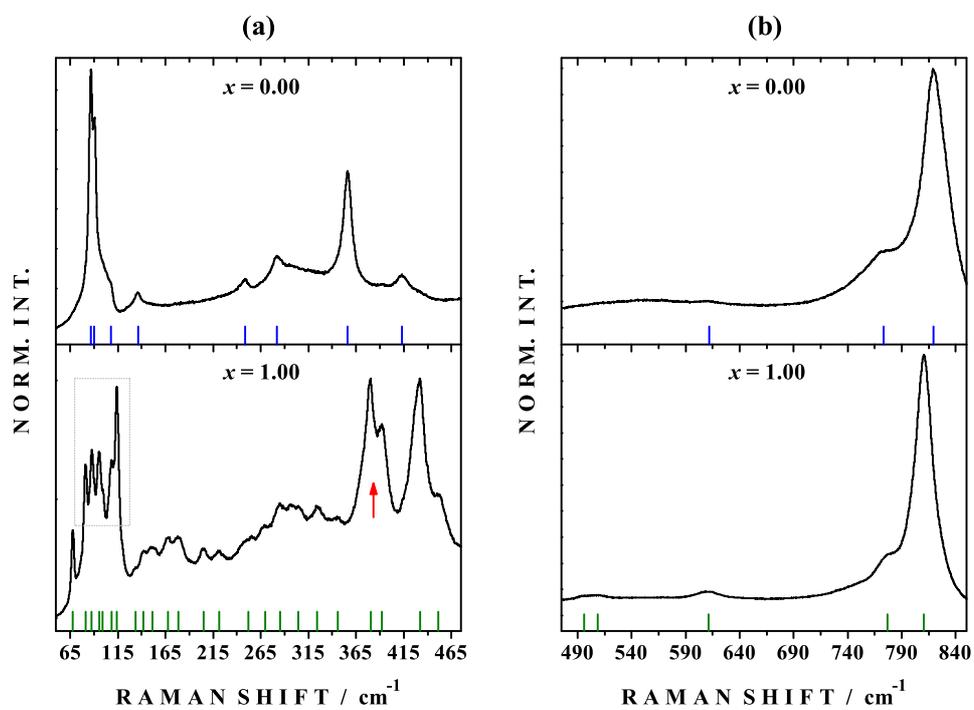

Figure 5:

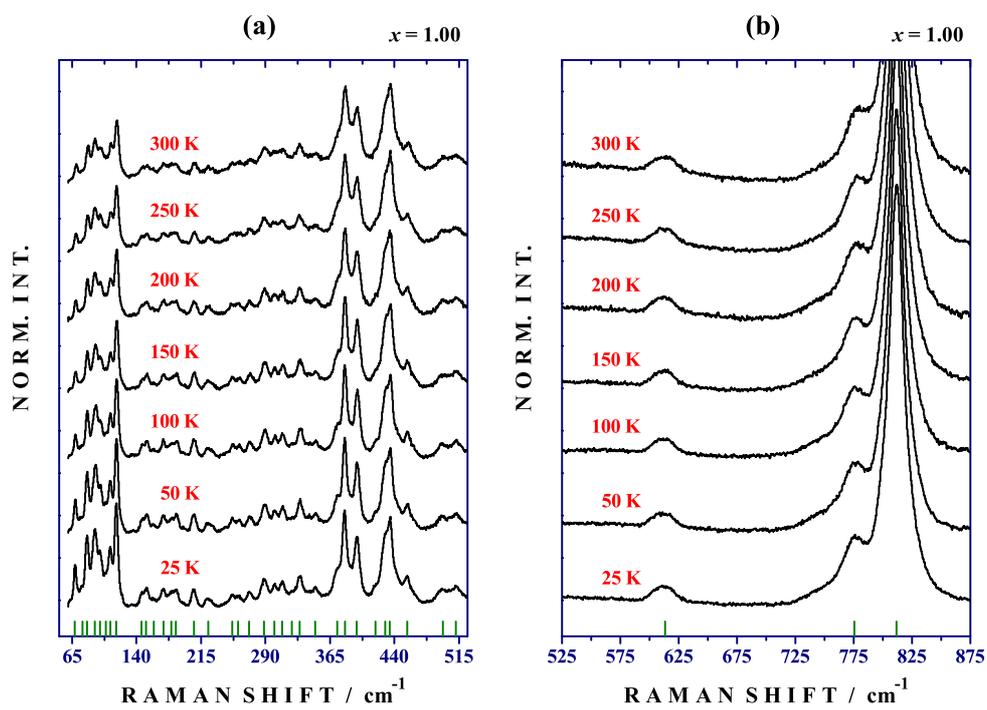

Table 1:

| Ion | Wyckoff site | Symmetry | Irreducible representation |
|---|---|---|---|
| **Trigonal:** $D_{3d}^3$; $P\text{-}3m1$; #164; $a^0a^0a^0$ | | | |
| $Ba^{2+}$ | $1b$ | $D_{3d}$ | $A_{2u} \oplus E_u$ |
| $Ba^{2+}$ | $2d$ | $C_{3v}$ | $A_{1g} \oplus A_{2u} \oplus E_g \oplus E_u$ |
| $Ca^{2+}$ | $1a$ | $D_{3d}$ | $A_{2u} \oplus E_u$ |
| $Nb^{5+}$ | $2d$ | $C_{3v}$ | $A_{1g} \oplus A_{2u} \oplus E_g \oplus E_u$ |
| $O^{2-}$ | $3f$ | $C_{2h}$ | $A_{1u} \oplus 2A_{2u} \oplus 3E_u$ |
| $O^{2-}$ | $6i$ | $C_S$ | $2A_{1g} \oplus A_{1u} \oplus A_{2g} \oplus 2A_{2u} \oplus 3E_g \oplus 3E_u$ |
| $\Gamma_T$ | $4A_{1g} \oplus A_{2g} \oplus 5E_g \oplus 2A_{1u} \oplus 8A_{2u} \oplus 10E_u$ | | |
| $\Gamma_{Ac}$ | $A_{2u} \oplus E_u$ | | |
| $\Gamma_{Si}$ | $A_{2g} \oplus 2A_{1u}$ | | |
| $\Gamma_{IR}$ | $7A_{2u} \oplus 9E_u$ | | |
| $\Gamma_R$ | $4A_{1g} \oplus 5E_g$ | | |
| **Trigonal:** $D_{3d}^4$; $P\text{-}3c1$; #165; $a^-a^-a^-$ | | | |
| $Sr^{2+}$ | $2a$ | $D_3$ | $A_{2g} \oplus A_{2u} \oplus E_g \oplus E_u$ |
| $Sr^{2+}$ | $4d$ | $C_3$ | $A_{1g} \oplus A_{1u} \oplus A_{2g} \oplus A_{2u} \oplus 2E_g \oplus 2E_u$ |
| $Ca^{2+}$ | $2b$ | $S_6$ | $A_{1u} \oplus A_{2u} \oplus 2E_u$ |
| $Nb^{5+}$ | $4d$ | $C_3$ | $A_{1g} \oplus A_{1u} \oplus A_{2g} \oplus A_{2u} \oplus 2E_g \oplus 2E_u$ |
| $O^{2-}$ | $6f$ | $C_2$ | $A_{1g} \oplus A_{1u} \oplus 2A_{2g} \oplus 2A_{2u} \oplus 3E_g \oplus 3E_u$ |
| $O^{2-}$ | $12g$ | $C_1$ | $3A_{1g} \oplus 3A_{1u} \oplus 3A_{2g} \oplus 3A_{2u} \oplus 6E_g \oplus 6E_u$ |
| $\Gamma_T$ | $6A_{1g} \oplus 8A_{2g} \oplus 14E_g \oplus 7A_{1u} \oplus 9A_{2u} \oplus 16E_u$ | | |
| $\Gamma_{Ac}$ | $A_{2u} \oplus E_u$ | | |
| $\Gamma_{Si}$ | $8A_{2g} \oplus 7A_{1u}$ | | |
| $\Gamma_{IR}$ | $8A_{2u} \oplus 15E_u$ | | |
| $\Gamma_R$ | $6A_{1g} \oplus 14E_g$ | | |
| **Monoclinic:** $C_{2h}^3$; $A2/m$; #12; $a^0b^-b^-$ | | | |
| $Sr^{2+}$ | $4i$ | $C_S$ | $2A_g \oplus A_u \oplus B_g \oplus 2B_u$ |
| $Sr^{2+}$ | $4i$ | $C_S$ | $2A_g \oplus A_u \oplus B_g \oplus 2B_u$ |
| $Sr^{2+}$ | $4i$ | $C_S$ | $2A_g \oplus A_u \oplus B_g \oplus 2B_u$ |
| $Ca^{2+}$ | $2a$ | $C_{2h}$ | $A_u \oplus 2B_u$ |
| $Ca^{2+}$ | $2d$ | $C_{2h}$ | $A_u \oplus 2B_u$ |
| $Nb^{5+}$ | $4i$ | $C_S$ | $2A_g \oplus A_u \oplus B_g \oplus 2B_u$ |
| $Nb^{5+}$ | $4i$ | $C_S$ | $2A_g \oplus A_u \oplus B_g \oplus 2B_u$ |
| $O^{2-}$ | $4i$ | $C_S$ | $2A_g \oplus A_u \oplus B_g \oplus 2B_u$ |
| $O^{2-}$ | $4i$ | $C_S$ | $2A_g \oplus A_u \oplus B_g \oplus 2B_u$ |
| $O^{2-}$ | $4i$ | $C_S$ | $2A_g \oplus A_u \oplus B_g \oplus 2B_u$ |
| $O^{2-}$ | $8j$ | $C_1$ | $3A_g \oplus 3A_u \oplus 3B_g \oplus 3B_u$ |
| $O^{2-}$ | $8j$ | $C_1$ | $3A_g \oplus 3A_u \oplus 3B_g \oplus 3B_u$ |
| $O^{2-}$ | $8j$ | $C_1$ | $3A_g \oplus 3A_u \oplus 3B_g \oplus 3B_u$ |
| $\Gamma_T$ | $25A_g \oplus 17B_g \oplus 19A_u \oplus 29B_u$ | | |
| $\Gamma_{Ac}$ | $A_u \oplus 2B_u$ | | |
| $\Gamma_{Si}$ | $0$ | | |
| $\Gamma_{IR}$ | $18A_u \oplus 27B_u$ | | |
| $\Gamma_R$ | $25A_g \oplus 17B_g$ | | |